\documentclass[twocolumn,prd,superscriptaddress,preprintnumbers,nofootinbib]{revtex4}[11pt]
\pdfoutput=1
\usepackage{amsmath,amssymb,graphicx}
\graphicspath{{figs/}}
\usepackage{epsf,verbatim}
\usepackage{hyperref}
\usepackage{comment}
\usepackage{color}
\usepackage{slashed}
\usepackage{subfigure}
\usepackage[usenames,dvipsnames]{xcolor}
\usepackage{comment}
\usepackage{mdwlist, paralist}
\usepackage{rotating}
\usepackage{multirow}

\usepackage[utf8]{inputenc}

\newcommand{\eg}{{\it e.g. }}
\newcommand{\Br}{{\mathrm{Br}}}

\def\to{\rightarrow}

\def\TeV{~{\mbox{TeV}}}
\def\abi{~{\mbox{ab}^{-1}}}
\def\fbi{~{\mbox{fb}^{-1}}}
\def\fb{~{\mbox{fb}}}
\def\GeV{~{\mbox{GeV}}}

\emergencystretch=\maxdimen
\begin{document}

\title{Invisible Higgs Decay at the LHeC}

\author{Yi-Lei Tang}
\thanks{tangyilei15@pku.edu.cn}
\affiliation{Center for High Energy Physics, Peking University, Beijing 100871, China}

\author{Chen Zhang}
\thanks{larry@pku.edu.cn}
\affiliation{Institute of Theoretical Physics $\&$ State Key Laboratory of Nuclear Physics and Technology, Peking University, Beijing 100871, China}

\author{Shou-hua Zhu}
\thanks{shzhu@pku.edu.cn}
\affiliation{Center for High Energy Physics, Peking University, Beijing 100871, China}
\affiliation{Institute of Theoretical Physics $\&$ State Key Laboratory of Nuclear Physics and Technology, Peking University, Beijing 100871, China}
\affiliation{Collaborative Innovation Center of Quantum Matter, Beijing 100871, China}

\begin{abstract}
The possibility that the 125 GeV Higgs boson may decay into invisible
non-standard-model (non-SM) particles is theoretically and phenomenologically
intriguing. In this letter we investigate the sensitivity of the Large Hadron
Electron Collider (LHeC) to an invisibly decaying Higgs, in its proposed high
luminosity running mode. We focus on the neutral current Higgs
production channel which offers more kinematical handles than its charged
current counterpart. The signal contains one electron, one jet and large missing
energy. With a cut-based parton level analysis, we estimate that if the $hZZ$ coupling
is at its standard model (SM) value, then assuming an integrated luminosity of
$1\,\mbox{ab}^{-1}$ the LHeC with the proposed 60 GeV electron beam (with
$-0.9$ polarization) and 7 TeV proton beam is capable of probing
$\mathrm{Br}(h\to\slashed{E}_T)=6\%$ at $2\sigma$ level. Good lepton veto
performance (especially hadronic $\tau$ veto) in the forward region is crucial to
the suppression of the dominant $Wje$ background. We also explicitly point out the
important role that may be played by the LHeC in probing a wide class of exotic
Higgs decay processes and emphasize the general function of lepton-hadron
colliders in precision study of new resonances after their discovery in
hadron-hadron collisions.
\end{abstract}

\maketitle

\setcounter{equation}{0} \setcounter{footnote}{0}

\section{Introduction}
After the discovery of the 125 GeV Higgs boson~\cite{Aad:2012tfa,Chatrchyan:2012ufa},
naturally the next step is measuring its properties as accurately as possible, which
tests the standard model (SM) in its most elusive sector and may hopefully reveal
its connection to physics beyond the standard model (BSM). So far, determination of
the Higgs boson spin and parity and
measurements of the Higgs signal strength in various production and decay
channels have been carried out, all of which
turned out to be consistent with SM predictions. It is worth noting that besides the
decay modes which have promising observability in SM, attention has also been paid
to interesting rare (such as flavor-changing~\cite{Khachatryan:2015kon,Mao:2015hwa})
or exotic~\cite{Curtin:2013fra} decay modes which is predicted to be negligibly
small in SM. These modes may easily get enhanced in various BSM theories, and with
the potential large number of Higgs bosons expected to be produced at the LHC and
future colliders, may even surprise us with a spectacular
discovery~\cite{Curtin:2013fra}.

One of the most interesting exotic Higgs decay channel is Higgs decaying into
invisible (or undetectable) non-SM particles~\cite{Shrock:1982kd,Martin:1999qf}.
Already long before the Higgs boson discovery, search of this mode has drawn a lot
of attention~\cite{Gunion:1993jf,Choudhury:1993hv,Eboli:2000ze,
Godbole:2003it,Davoudiasl:2004aj,Zhu:2005hv}. Near and after the Higgs boson
discovery, constraints has been put on
the invisible Higgs decay branching fraction via both Higgs signal strength
measurements and direct searches. The most stringent limit from direct search now
comes from the ATLAS search for an invisibly decaying Higgs in the vector boson
fusion (VBF) channel~\cite{Aad:2015txa}, which constrains the Higgs invisible
branching fraction to be less than 29\% at 95\% confidence level. The importance
of this exotic Higgs decay channel however cannot be overemphasized because it may
shed light on the link between Higgs boson and the dark matter (DM), whose
existence has been established via its gravitational effects. It is tempting to
speculate about their connection since there is no evidence of DM interacting
non-gravitationally with other SM particles which have been found for a long time
and also the Higgs is one of the only few portals by which SM particles are able
to interact with singlet hidden sector matter via (super-)renormalizable
Lagrangians. Indeed the situation that the Higgs (or the Higgs sector)
interacting with DM (or a dark sector) occurs in many extensions of the SM, for
example in models which aim to solve the hierarchy problem, such as supersymmetry
(SUSY)~\cite{Griest:1987qv,Gunion:1988yc,Djouadi:1996mj,Djouadi:1997gw,Cao:2012im},
composite Higgs~\cite{Fonseca:2015gva}, extra dimensions~\cite{Giudice:2000av},
Little Higgs~\cite{Hundi:2006rh}, Twin Higgs~\cite{Craig:2013fga,Liu:2014rqa},
in simple dark matter models~\cite{Burgess:2000yq,Gopalakrishna:2009yz,
Feng:2014vea}, and in Higgs portal models~\cite{Englert:2011yb,Mambrini:2011ik,
Djouadi:2011aa,He:2011gc,Djouadi:2012zc,Anchordoqui:2013bfa,Baek:2014jga}. If the dark matter
(or dark sector) particle
is sufficiently light, then the invisible Higgs decay thereof can naturally reach
a detectable branching fraction. Invisible Higgs decay is also an important
signature of some majoron and neutrino mass models~\cite{Joshipura:1992hp,
Ghosh:2011qc,Banerjee:2013fga,Bonilla:2015uwa,Seto:2015rma}. In fact the
collider search for invisible Higgs decay has become and will remain
an important constraint on wide classes of BSM scenarios. It is thus highly
motivated to investigate all sensitive search strategies within the possibly
available accelerator and detector designs.

\begin{figure}[ht]
\includegraphics[width=2.2in]{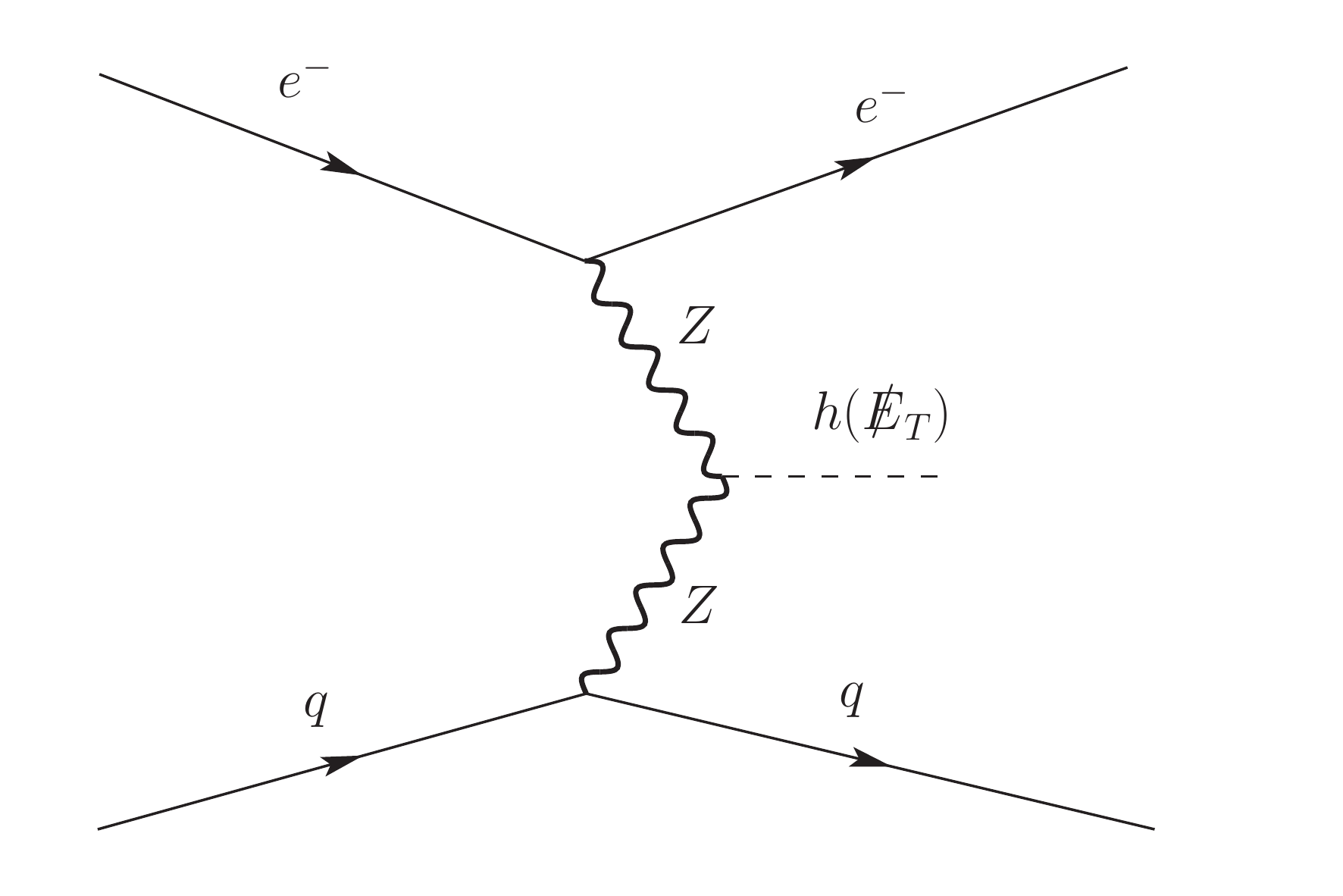}
\caption{Feynman diagram of the NC production of an invisible Higgs at the LHeC.\label{fig:h2INVSig}}
\end{figure}

At the LHC, it has been recognized that the VBF and ZH associated production
channels will provide the most sensitive probe on an invisibly decaying Higgs
in the long run~\cite{Aad:2009wy,Bai:2011wz,Ghosh:2012ep,Bernaciak:2014pna}.
At lepton colliders such as the International Linear
Collider (ILC), Future Circular Collider (FCC-ee) or the Circular Electron
Positron Collider (CEPC), sensitivity to $\Br(h\to\slashed{E}_T)<1\%$ can
easily be gained due to the much cleaner collider environment and the
availability of mass recoil methods~\cite{Baer:2013cma,cepc:2015pcdr}.
However it is still helpful to investigate whether other options exist and
may help to provide useful information on our understanding of physics
behind the scene.

In this letter, we investigate the possibility of utilizing the Large Hadron
Electron Collider (LHeC)~\cite{AbelleiraFernandez:2012cc} with its recently
proposed and discussed high luminosity
run~\cite{Zimmermann:2013aga,Bruning:2015acc,Onofrio:2015phy} to
probe an invisibly decaying Higgs. The LHeC plans to collide a 60 GeV
electron beam with the 7 TeV proton beam in the LHC ring and is designed to
run synchronously with the High Luminosity Large Hadron Collider (HL-LHC). It was originally designed to deliver an
integrated luminosity of $100\fbi$. With the potential role of the LHeC in
precision Higgs physics being noticed, recently there has been proposals
and discussion on the collider's luminosity upgrade which is designed to
deliver an integrated luminosity of up to
$1\,\abi$~\cite{Zimmermann:2013aga,Bruning:2015acc,Onofrio:2015phy}.
With such conditions the LHeC indeed becomes a Higgs boson factory and
will offer exciting opportunities in precision Higgs studies, especially
with respect to exotic Higgs decays. We note that there has been quite a
few studies on Higgs boson physics at the LHeC~\cite{Han:2009pe,
Zhe:2011yr,Biswal:2012mp,Senol:2012fc,Cakir:2013bxa,Yue:2015dia,
Liu:2015kkp,Kumar:2015tua,Kumar:2015jna}. The possibility of using the
LHeC to study BSM Higgs decays has been mentioned
in~\cite{Onofrio:2015phy}.

At the LHeC the Higgs boson is produced via two major channels:
charged current (CC) and neutral current (NC). In CC production, the Higgs
is produced via $WW$ fusion and has a larger cross section. However, when
searching for an invisible Higgs, $WW$ fusion production results in
mono-jet plus missing energy, which accidentally coincides with the CC
deeply inelastic scattering (DIS) background. Moreover, the lack of
kinematic handles in the final state renders this signal channel even
more difficult to distinguish from its major background. Therefore
in this letter we focus on NC production which results in one electron,
one jet and large missing energy in the final state (see
FIG. ~\ref{fig:h2INVSig} for its Feynman diagram). In the next section
we present a cut-based parton level analysis of the relevant signal and
backgrounds and derive an estimation on the LHeC sensitivity to invisible
Higgs decay branching fraction with its high luminosity mode. The main
conclusion is that the LHeC in its high luminosity mode has the potential
to probe $\mathrm{Br}(h\to\slashed{E}_T)=6\%$ at $2\sigma$ level
(assuming a SM $hZZ$ coupling), when only statistical uncertainty
is taken into account. We will also
discuss about crucial assumptions made in our analysis and comment on
possible future improvement and promising potential of lepton-hadron
colliders in precision studies of new resonances (including the study
of various exotic Higgs decays) in the last section before we conclude
the letter. We emphasize that an electron-proton collider with even
higher beam energies (\eg $E_e=120\GeV,E_p=50\TeV$) may has better
sensitivity to the invisible Higgs decay (and other exotic Higgs
decays), which is interesting and worth pursuing in its own right but
beyond the scope of this letter.

\begin{figure*}[ht]
\includegraphics[width=2.2in]{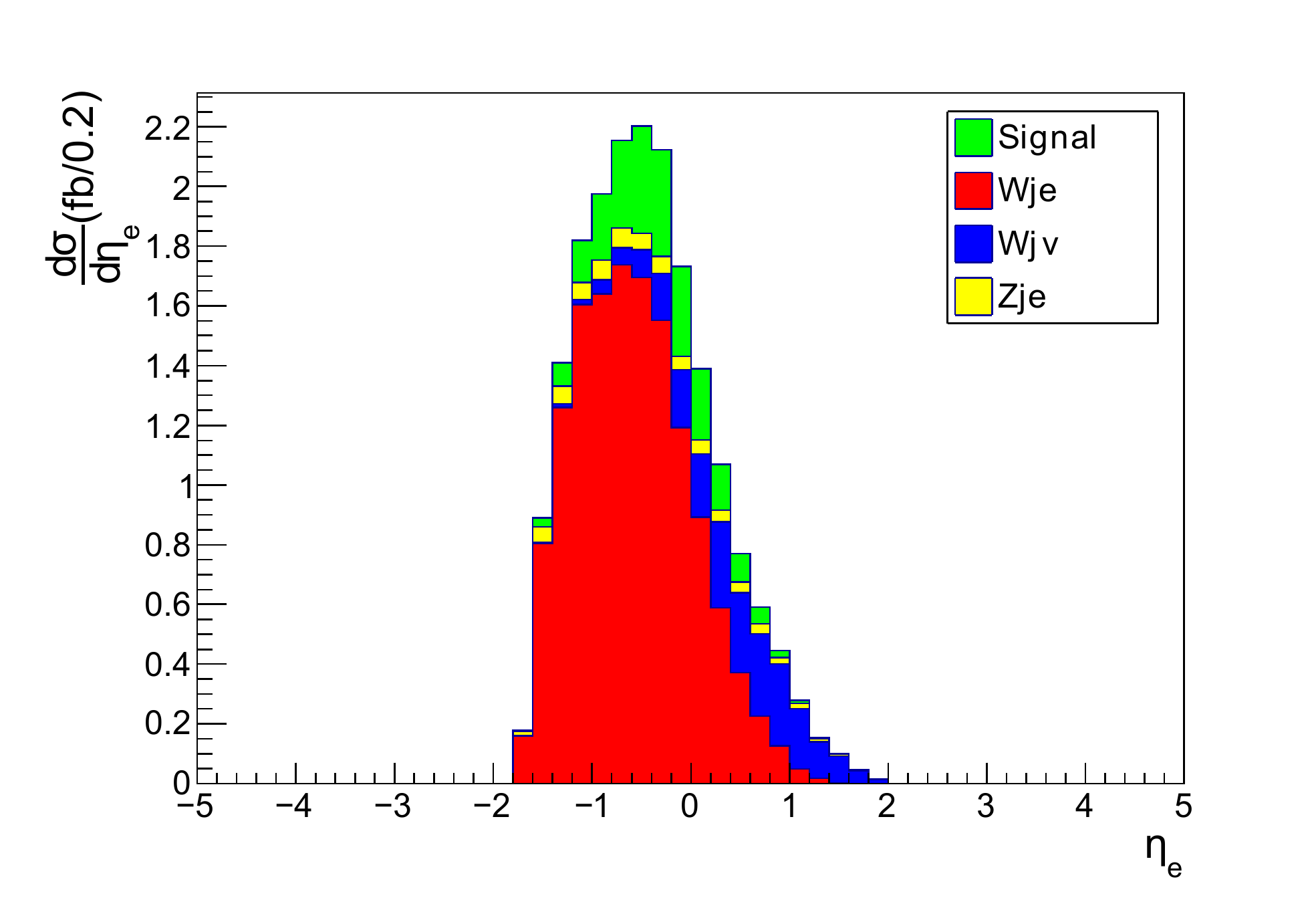}
\includegraphics[width=2.2in]{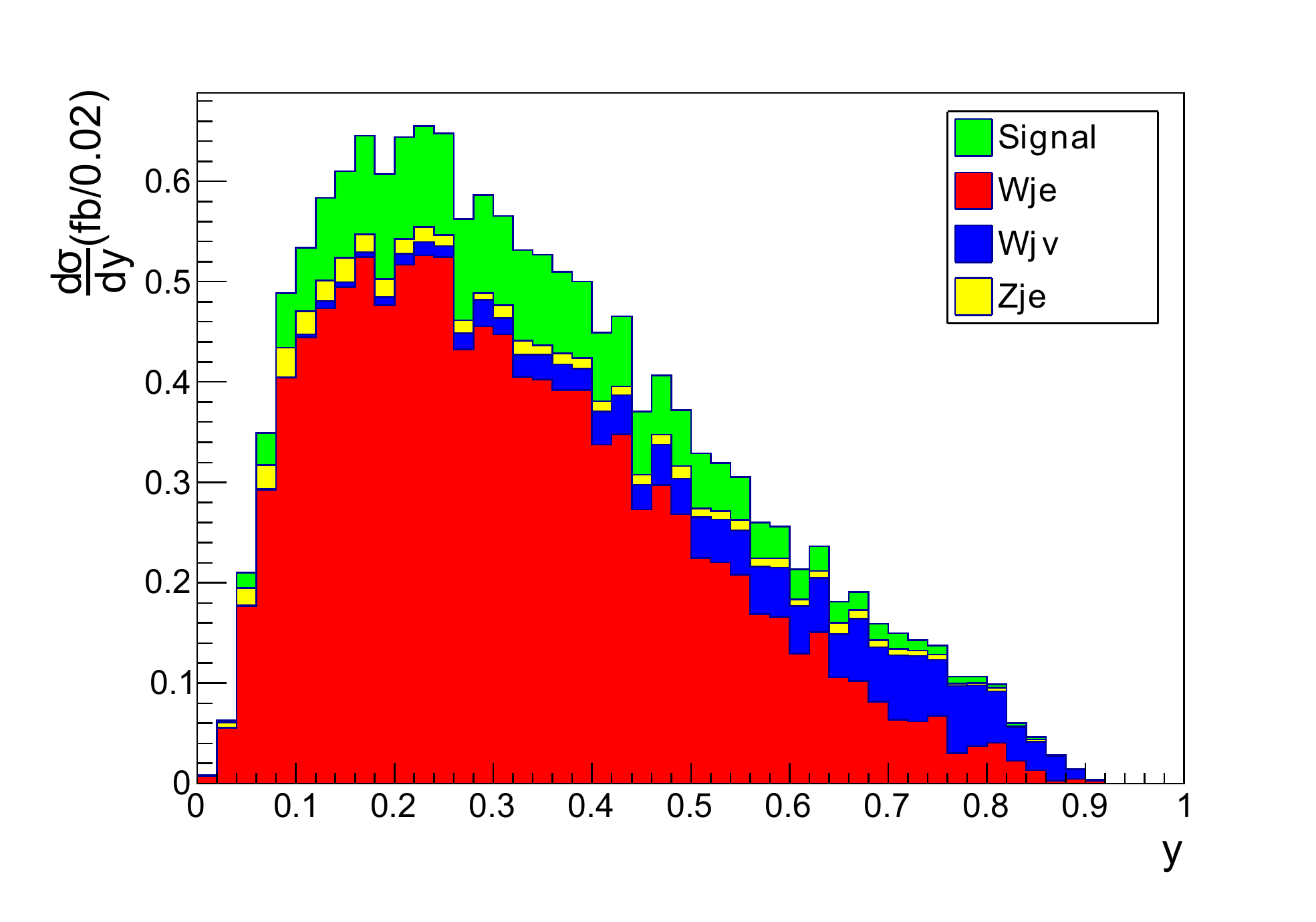}
\includegraphics[width=2.2in]{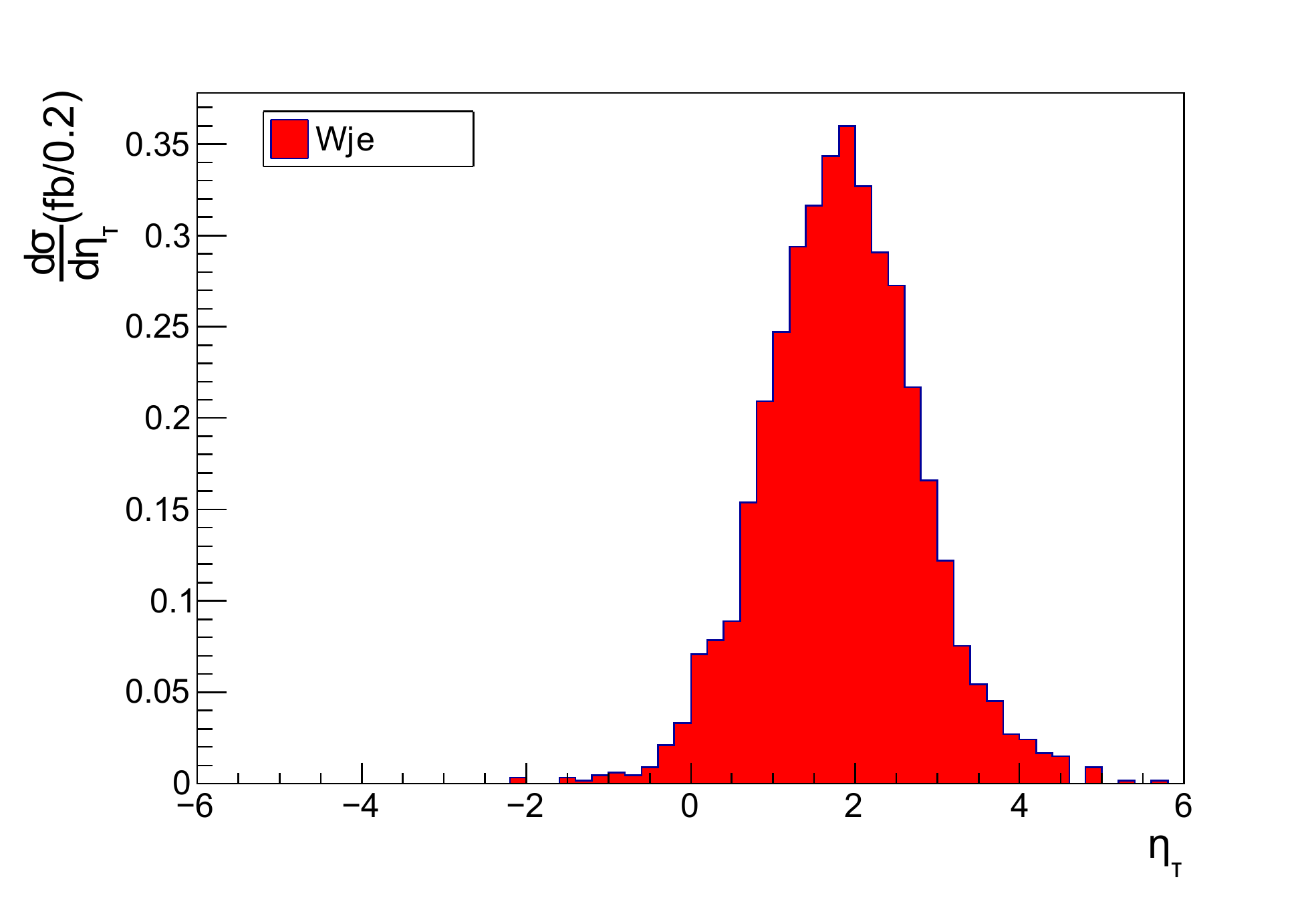}
\caption{\label{fig:distribution}Left: $\eta_e$ distribution of the signal and major backgrounds just before the $\eta_e$ cut. Middle: $y$ distribution of the
signal and major backgrounds just before the $y$ cut. Right: $\tau$ lepton pseudorapidity distribution of the $Wje (W\to\tau\nu)$ background just before the
lepton veto.}
\end{figure*}

\section{Collider Sensitivity}
\subsection{Signal and Backgrounds}
We take Higgs production at the LHeC through $ZZ$ fusion as our signal
process. Its cross section depends on the $hZZ$ coupling in the model
considered. We use $\kappa_Z$ to denote the $hZZ$ coupling strength
relative to its SM value. Then we define
\begin{align}
C_{\text{MET}}^2=\kappa_{Z}^2\times\Br(h\rightarrow\slashed{E}_T)
\label{eqn:cmet2}
\end{align}
with which we are able to present the sensitivity results conveniently.
The SM process $h\to ZZ^\ast\to 4\nu$ has an extremely small
branching ratio and will neither be included in the signal nor
backgrounds.

The main irreducible backgrounds include
\begin{eqnarray}
& & p + e^- \rightarrow W^- + j + \nu_e, (Wj\nu)\\
& & p + e^- \rightarrow Z + j + e^-, (Zje)
\end{eqnarray}
which result in
one electron, one jet and missing transverse energy via $W\to e\nu$
and $Z\to \nu\nu$ respectively.
Photoproduction of $W+j$ is also an irreducible background
if the $W$ boson decays to an electron. Although its cross section is
initially very large, it is found to be negligible after all selection
cuts described below, due to its distinct kinematic features. We do not
expect the $W+j$ production via resolved photons contributes
sizably to the total background because we require large missing energy
in the event which should boost the $W$ boson to the kinematic regime
where the resolved photon contribution is expected to be
small~\cite{Diener:2002if}.

We note that these irreducible backgrounds do not contain strong
coupling at leading order, which is different from the VBF search
for an invisible Higgs encountered at the LHC. At the LHC, the VBF
search of an invisible Higgs boson has important $Vjj(V=W,Z)$
backgrounds which contain the process of QCD dijet production with
a weak boson radiated from one of the quark line. However at the
LHeC the corresponding process is of purely electroweak nature
which is one of the attractive features of a lepton-hadron
collider machine.

There are also reducible backgrounds which come from a variety
of sources. Anti-top production in which $\bar{t}$ decays to $\bar{b}+e^{-}+\bar{\nu_e}$
constitutes a background because $b$ anti-tagging cannot be expected
to be fully efficient. However this background also turns out to be
negligible after all selection cuts below. A threatening reducible
background is
\begin{eqnarray}
& & p + e^- \rightarrow W^{\pm} + j + e^-, (Wje)
\end{eqnarray}
in which the $W$ boson decays to $l\nu\,(l=e,\mu,\tau)$ and the
charged lepton $(e,\mu,\tau)$ from $W$ decay falls out of detector
acceptance or fails to be reconstructed and identified. In fact
this background turns out to be dominant after all selection cuts.

$e+\text{multijet}$ production is a reducible background in which
missing energy comes from jet energy mismeasurement. We do not
simulate this background but its contribution is expected to be
negligible after several demanding cuts required in the analysis,
especially $\slashed{E}_T>70\GeV$ and the missing energy
isolation cut $I\equiv\Delta\phi_{\slashed{E}_T,j}>1\,\mathrm{rad}$.
One further reducible background is CC $jj\nu$ production in
which one jet is misidentified as an electron. In the following
we simply assume a competent detector performance and drop this
background from the analysis.

\subsection{Analysis and Results}

\begin{table*}[t!]
\begin{tabular}{|c|c|c|c|c|c|c|c|c|}
\hline
Cross Section (fb) & Basic Cuts & $\slashed{E}_T>70\GeV$ & $I>1$ & $\eta_j-\eta_e>3.0$ & $\Delta\phi_{ej}<1.2$ & $\eta_e\in[-1.2, 0.6]$ & $y\in [0.06, 0.5]$ & Lepton Veto \\
\hline
Signal ($C_{\text{MET}}^2=1$) & 16.1 & 8.80 & 8.23 & 4.68 & 2.37 & 2.16 & 1.77 & 1.77 \\
\hline
$Wje$ & 816 & 158 & 143 & 51.7 & 13.9 & 11.3 & 9.13 & 1.96 \\
\hline
$Wj\nu$ & 192 & 102 & 101 & 5.68 & 2.36 & 1.33 & 0.387 & 0.387 \\
\hline
$Zje$ & 42.7 & 13.8 & 12.1 & 1.64 & 0.683 & 0.464 & 0.326 & 0.326 \\
\hline
\end{tabular}
\caption{The cross section (in unit of fb) of the signal and major backgrounds after application of each cut in the corresponding column. Other backgrounds contribute
less than $0.1\fb$ in total after all cuts and are not displayed in the table.
\label{table:cutflow}}
\end{table*}

We generate the signal and background samples at leading order with
MadGraph5\_aMC@NLO~\cite{Alwall:2014hca}. The collider parameters
are taken to be $E_e=60\GeV, E_p=7\TeV$ with the electron beam
being $-0.9$ polarized, i.e. $90\%$ left-handed. These parameter
choices including polarization are in accordance with the LHeC Higgs
factory parameters presented in~\cite{Zimmermann:2013aga}.
The Higgs boson mass is taken to be 125 GeV. The parton distribution
function used is NNPDF2.3 at leading order~\cite{Ball:2013hta}.
We take the renormalization and factorization scale to be the
$Z$ boson mass. For all the signal and background considered, K-factors are taken
to be 1~\cite{Stelzer:1997ns,deBoer:2007zz,Jager:2010zm}. We perform a
parton level analysis with detector resolution taken into account
by the jet and lepton energy resolution formula
$\frac{\sigma_E}{E}=\frac{\alpha}{\sqrt{E}}\oplus\beta$
where for jet energy smearing $\alpha=0.6\GeV^{1/2},\beta=0.03$
and for lepton energy smearing $\alpha=0.05\GeV^{1/2},\beta=0.0055$.
Event analysis is performed with the help of
MadAnalysis 5~\cite{Conte:2012fm}. Whenever needed, the expected
statistical significance $Z$ is calculated according to the
formula $Z=\sqrt{2((S+B)\ln(1+S/B)-S)}$ ~\cite{Cowan:2010js}
where $S$ and $B$ denote the expected signal and background event
number, respectively.

As to signal and background analysis, we require at least one
electron and at least one jet in the final state. All the signal
and background samples are required to pass the following basic
cuts:
\begin{align}
& p_{Tj}>20\GeV, |\eta_j|<5.0, \nonumber \\
& p_{Tl}>20\GeV, |\eta_l|<5.0, \Delta R_{jl}>0.4
\label{eqn:bcut}
\end{align}
Then we impose the following sequence of selection cuts to
further discriminate between signal and background:
\begin{enumerate}
\item $\slashed{E}_T>70\GeV$.
\item Missing energy isolation: $I>1\,\mathrm{rad}$.
\item Pseudorapidity gap of the jet and the electron satisfies $\eta_{j} - \eta_{e}>3.0$.
\item The azimuthal angle difference of the electron and the jet satisfies $\Delta\phi_{ej}\equiv|\phi_j-\phi_e|<1.2$.
\item The pseudorapidity of the electron satisfies $\eta_e\in[-1.2, 0.6]$.
\item Inelasticity cut: the inelasticity variable $y$ is defined as
$y = \frac{p_1 \cdot ( k_1 - k_2 )}{p_1 \cdot k_1}$, where $p_1$ is
the 4-momenta of the initial proton, $k_1$ is the 4-momenta of the
initial electron, $k_2$ is the 4-momenta of the out-going electron. Then we require $y \in [0.06, 0.5]$.
\item Lepton veto: additional electron, muon or tagged hadronic $\tau$ are vetoed. (See text for detail.)
\end{enumerate}
We assume additional electrons satisfying $p_T>7\GeV$ and $|\eta|<5.0$ and muons
satisfying $p_T>5\GeV$ and $|\eta|<5.0$ can all be vetoed. As to the
$\tau$ decay, we adopt the collinear approximation in which we simply assume on average
the visible electron or muon from $\tau$ decay carries $1/3$ of the
parent $\tau$ momentum and the visible part of a hadronically
decaying $\tau$ carries $1/2$ of the parent $\tau$
momentum. We consider a $70\%$ tagging efficiency~\cite{Aad:2014rga} for a hadronically
decaying $\tau$ for the veto purpose, if the $\tau$ lepton satisfies
$p_{T,\tau_{\text{had-vis}}}>20\GeV$ and $|\eta|<5.0$ ($p_{T,\tau_{\text{had-vis}}}$
denotes the transverse momentum of the visible part of the
hadronically decaying $\tau$). We note
that we have allowed the lepton veto capability to extend to $|\eta_{\text{max}}|=5.0$,
in contrast to the commonly assumed $|\eta_{\text{max}}|=2.5$ assumed in the usual LHC
analysis. This is due to the expected very large pseudorapidity coverage of
the LHeC tracking detector and muon
detector~\cite{AbelleiraFernandez:2012cc,Gaddi:2015det}.

\begin{figure}[ht]
\includegraphics[width=2.5in]{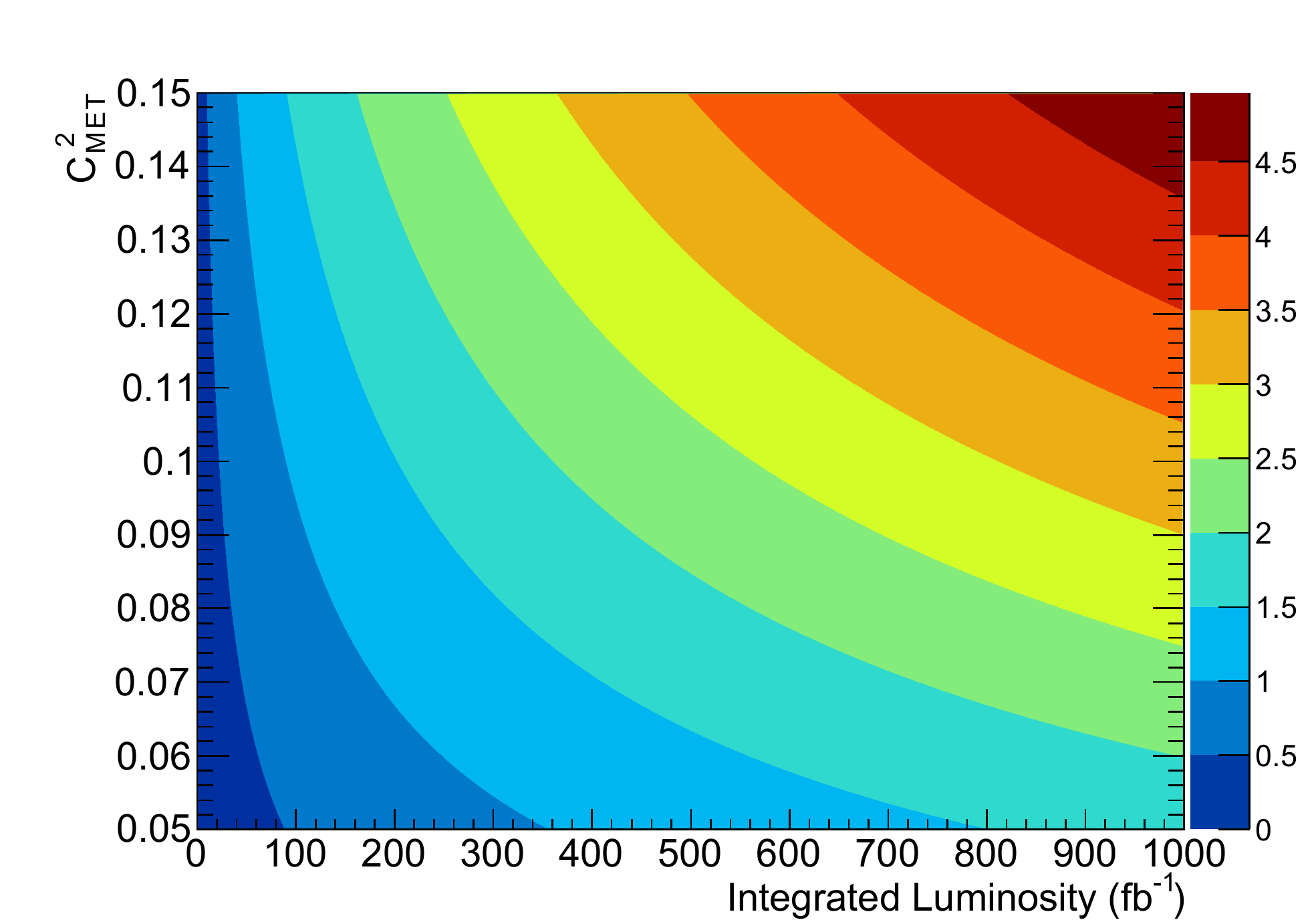}
\caption{The expected significance contour for an invisible Higgs at the LHeC.
The colors indicate the value of the expected statistical significance
with the correspondence displayed by the scale on the right.\label{fig:significance}}
\end{figure}

In the sequence of cuts listed above, $\slashed{E}_T>70\GeV$ and the missing energy
isolation requirement will significantly suppress the $e+\text{multijet}$
background. When calculating the missing energy the electron and hadronic
$\tau$ which satisfy $|\eta|<5.0$ but fail to be identified are counted in
the $p_T$ balance while muons which fail to be identified are always excluded
in the $p_T$ balance. The pseudorapidity gap requirement, azimuthal
angle difference cut are analogous to $|\eta_{j_1}-\eta_{j_2}|$
and $\phi_{jj}$ cuts employed in the LHC VBF search for an invisible
Higgs boson~\cite{Eboli:2000ze}. They are very effective in reducing
all three major backgrounds. Then we try simple kinematic
variables like the electron pseudorapidity
$\eta_e$ and inelasticity $y$ to further enhance the statistical
significance. To motivate those cuts beyond the counterparts of the
usual ones employed in the VBF search for an invisible Higgs at the LHC,
we plot the $\eta_e,y$ distribution of the signal and major backgrounds in
FIG. ~\ref{fig:distribution} (left and middle) just before applying the correspongding
cuts. The signal and background cross sections after each cut are listed in
Table~\ref{table:cutflow}, in which the signal cross section is calculated
assuming $C_{\text{MET}}^2=1$. We note that if we target
$C_{\text{MET}}^2=0.06$, we will get an expected signal cross
section of 0.106fb and total background cross section 2.761fb (we have
included here about $0.1\fb$ contribution from other minor backgrounds),
thus $S/B\approx3.8\%$, and with an integrated luminosity of
$1\,\abi$ the expected statistical significance will reach
$Z=2.00$. We also plot the significance contour for a targeted range
of $C_{\text{MET}}^2$ and the luminosity parameter in FIG. ~\ref{fig:significance}.

\section{Discussion and Conclusion}
In this letter we have studied at the parton level the possibility
of using the LHeC to search for an invisibly decaying Higgs boson
and find that the LHeC has promising potential to discover or
constrain this important exotic Higgs decay mode.
The $\eta_e$ cut and the inelasticity cut are found to be very effective in
suppressing the $Wj\nu$ background which has no counterpart in
the VBF search for an invisible Higgs at the LHC. After all selection
cuts the largest background turns out to be the $Wje$ process
in which the charged leton (especially $\tau$) from $W$ boson decay
fails to be identified. In fact this $Wje$ process finally
constitutes about $70\%$ of the total background. We take advantage of the expected large
acceptance of the LHeC tracking detector and muon detector, with
which the lepton veto is able to remove nearly $80\%$ of the $Wje$
background. Especially we find that the lepton veto capability in
the forward region $\eta\in[2.5,4.0]$ is essential. To illustrate
this point we plot the pseudorapidity distribution of the $\tau$
lepton from the $W$ boson decay in the $Wje$ background
(FIG. ~\ref{fig:distribution} (right)). Due to lepton universality
this also represents the pseudorapidity distribution of the
electron/muon from $W$ boson decay in the $Wje$ background. From
the plot it is clear that the charged lepton from $W$ boson
decay in $Wje$ background are mostly distributed in
$\eta\in[0.0,4.0]$ and large portions of events still
reside in $\eta\in[2.5,4.0]$. If the lepton veto is only possible
in $\eta\in[-2.5,2.5]$, the total background will nearly double.
It is thus highly recommended that a good lepton veto capability
should be maintained in the forward region $\eta\in[2.5,4.0]$.

The sensitivity of the LHeC to an invisibly decaying Higgs
boson could be further enhanced via a multivariate analysis,
which is worth pursuing~\cite{Tang:2015mul} but beyond the
scope of the present paper. Compared with the concurrent
search at the HL-LHC, the invisible Higgs search at the
LHeC has the further advantage of not suffering from
pile-up, a crucial factor of which is commonly not taken
account sufficiently in the LHC analysis.
Of course both searches are worth exploiting and
being combined to produce the best sensitivity to the
invisible Higgs decay with the available LHC infrastructure.
Even if an excess of VBF $\text{dijet}+\slashed{E}_T$ events
is first observed at the HL-LHC, signals from additional
channels are still required to pin down the origin of the
$\slashed{E}_T$ signature. The LHeC search for an invisible
Higgs may play an important role in this process.

Our study clearly justifies a luminosity upgrade to
$1\,\abi$ for the LHeC to become a Higgs boson
factory~\cite{Zimmermann:2013aga} and demonstrates its huge
potential on study of exotic Higgs decays. Besides the
invisible Higgs decay, the LHeC is suited to the study of
those exotic Higgs decays which suffer from large backgrounds,
trigger or $p_T$ threshold problem at the (HL-)LHC such as
$h\to 4b$, $h\to 2b2\tau$, $h\to 4j$,
$h\to b\bar{b}+\slashed{E}_T$~\cite{Huang:2013ima},
$h\to\gamma+\slashed{E}_T$,
$h\to Z+\slashed{E}_T$~\cite{Liu:2013sx}. Work on these
directions is in progress~\cite{Zhang:2015ehd}. The
demonstration of the LHeC potential on studying exotic Higgs
decays reveals an important aspect of lepton-hadron colliders
with respect to precision study after the discovery of a new
resonance in hadron-hadron collisions, which has not been
unexpected since the early study of measuring the bottom
Yukawa coupling at the LHeC~\cite{Han:2009pe}. Although
usually the ideal precision measurement should finally be
achieved at a lepton collider, this most precise
measurement can only be reached with sufficient
center-of-mass energy available. Without the help of such a lepton
collider then the best use of a hadron beam can be made via
colliding it against a lepton beam to make foreseeable
precision studies, which may even unravel exciting
deviations from the SM within the shortest time.
\vfil

\subsection*{Acknowledgements}

We would like to thank Qing-Hong Cao, Tao Han, Qiang Li, Yan-Dong Liu,
Ying-Nan Mao, Jian-Ming Qian, Hui-Chao Song, Lian-Tao Wang
and Hao Zhang for helpful discussions. This work was supported
in part by the Natural Science Foundation of China
(Grants No. 11135003 and No. 11375014).


\bibliography{h2inv_v17}
\bibliographystyle{h-physrev}

\end{document}